\documentclass{elsart}

\usepackage{amssymb}
\usepackage{amsfonts}
\usepackage{amsmath}
\usepackage{color}

\newcommand{\beq}{\begin{equation}}
\newcommand{\eeq}{\end{equation}}
\newcommand{\bea}{\begin{eqnarray}}
\newcommand{\eea}{\end{eqnarray}}

\begin{document}

\begin{frontmatter}

\rightline{ICRR-Report-539,STUPP-09-203}

\title{The Universal Extra Dimensional Model with $S^2/Z_2$ extra-space }

\author[chuo]{Nobuhito Maru},
\ead{maru@phys.chuo-u.ac.jp}
\author[Saitama]{Takaaki Nomura \corauthref{cor}},
\corauth[cor]{Corresponding author. 
telephon number: 810488589102}
\ead{nomura@krishna.th.phy.saitama-u.ac.jp}
\author[Saitama]{Joe Sato} and
\ead{joe@phy.saitama-u.ac.jp}
\author[ICRR]{Masato Yamanaka}
\ead{yamanaka@icrr.u-tokyo.ac.jp}

\address[chuo]{
Department of Physics, Chuo University, Tokyo 112-8551, Japan}

\address[ICRR]{
University of Tokyo, Kashiwa, Chiba 277-8582, Japan}

\address[Saitama]{Department of Physics,
  Saitama University,
  Shimo-Okubo,
  Sakura-ku,
  Saitama 355-8570, Japan}

\begin{abstract}
We propose a new Universal Extra Dimensional model 
that is defined on 
a six-dimensional spacetime which has a two-sphere orbifold $S^2/Z_2$ as an extra-space.
We specify our model by choosing the gauge symmetry as SU(3)$\times$SU(2)$\times$U(1)$_Y \times$U(1)$_X$, 
introducing field contents in six-dimensions as their zero modes correspond to the Standard model particles,
and determining a boundary condition of these fields on orbifold $S^2/Z_2$.
A background gauge field 
that belongs to U(1)$_X$ is introduced there, which is necessary 
to obtain massless chiral fermions in four-dimensional spacetime.
We then analyze Kaluza-Klein(KK) mode expansion of the fields in our model and derive the mass spectrum of the 
KK particles.
We find that the lightest KK particles are the 1st KK particle of massless gauge bosons at tree level.  
We also discuss the KK parity of the KK modes in our model and confirm the stability of the 
lightest KK particle which is important for dark matter physics.

\end{abstract}


\end{frontmatter}

\section{Introduction}

The Standard Model(SM) has passed all the accelerator experiments. 
It is, however, not a satisfactory theory for all the physicists. 
There seem to be several problems, e.g. a large number of parameters 
(18 ! even without neutrino masses and lepton mixings), the hierarchy problems, and so on. 
Among them,
 the hierarchy problem strongly drives physicists to construct a model 
beyond the SM. 
We need to explain the stability of the weak scale to solve the problem. 
For 
such an explanation,
supersymmetry has been mostly employed 
and the consequence of these models are extensively explored. 
There are also other mechanisms, say, little higgs, extra dimensions,
 and so on. 
These have not been intensively studied 
compared to supersymmetric models. 
Furthermore there is still a room for new type of models.
Since the Large Hadron Collider experiment is about to operate, 
which will explore the physics at TeV scale, 
it is urgent to investigate all the possible models at that scale.

Among those mechanisms, 
the idea of Universal Extra Dimensional(UED) model is very interesting 
\cite{Appelquist:2000nn,Antoniadis:1990ew}. 
Indeed the minimal version of UED has recently been studied very much. 
It is a model with one extra dimension defined on an orbifold $S^1/Z_2$. 
This orbifold is given by identifying the extra spatial coordinate $y$ with $-y$ 
and hence there are fixed points $y=0,\pi$.
By this identification chiral fermions are obtained.
It is shown that this model is free from the current experimental constraints 
if the scale of extra dimension $1/R$, which is the inverse of the compactification radius $R$, 
is larger than 400 GeV \cite{Appelquist:2000nn},\cite{Agashe:2001ra}. 
The dark matter can be explained by the first or second Kaluza-Klein (KK) mode \cite{Cheng:2002ej}, 
which is often the first KK photon, and this model can be discriminated from other models \cite{Datta:2005zs}. 
This model also can give plausible explanations for SM neutrino masses which are 
embedded in extended models \cite{Matsumoto:2006bf}. 

The UED models with more than five dimensions have not been studied 
extensively
despite 
the fact that it can explain some problems in the SM. 
The
 six dimensional models are 
particularly interesting.  
It is known that the number of the generations of quarks and leptons is derived 
by anomaly cancellations \cite{Dobrescu:2001ae} and 
the proton stability is guaranteed by a discrete symmetry of a subgroup of 6D Lorentz symmetry 
\cite{Appelquist:2001mj}. 
Since the above UED model was proposed as a six dimensional model with extra dimensions of $T^2/Z_2$, 
it is very interesting to pursue six dimensional models with an alternative compactification.

As a physically intriguing example, 
there is a model with two dimensional compact space $S^2$, 
which has so far received a little attention 
(see for some works in this direction on the Einstein-Maxwell theory \cite{RandjbarDaemi:1982hi}, \cite{RandjbarDaemi:1983bw} 
and the gauge-Higgs unification \cite{Manton:1979kb}, \cite{Lim:2006bx}, \cite{Nomura:2008sx}).  
In models with two 
spheres, 
it is well known that fermions cannot be massless because of the positive curvature 
and hence they have a mass of O($1/R$) \cite{Lichnerowicz:1964zz,A.A.Abrikosov}.
We cannot overcome the theorem simply by the orbifolding of the extra spaces.
In another words, we have no massless fermion on the curved space with positive curvature, 
but we know a mechanism to obtain a massless fermion on that space
by introducing a nontrivial background gauge field \cite{Horvath:1977st,RandjbarDaemi:1982hi}.
The nontrivial background gauge field can cancel the spin connection term in the covariant derivative. 
As a result, a massless fermion naturally appears. 
Furthermore, we note that the background gauge field configuration is energetically favorable 
since the background gauge kinetic energy 
lowers a total energy. 
In order to realize chiral fermions, the orbifolding is required, for instance.

In this paper, we study a new type of UED with $S^2$ extra dimensions.
We will show that we can construct a model in six dimensions with $S^2/Z_2$ extra space. 
We extend the SM to the space, employ the method of background, and acquire chiral fermions.
Due to this orbifolding,
 all the bosons of the SM can be massless in the SU(2) limit. 
This means the lowest states are completely consistent with the SM as they should be. 
Furthermore, there are KK modes for each particle, 
and the lightest mode among them is stable due to the KK parity originated from the orbifolding.
Besides the complexity 
stemming from the structure of $S^2$ instead of $S^1$, 
the feature is quite similar up to the first KK mode. 
The difference appears from the second KK modes.

The paper is organized as follows. 
In Section \ref{S2UED}, we recapitulate a gauge theory defined on the six-dimensional spacetime which has 
two-sphere orbifold $S^2/Z_2$ extra-space and then specify our UED model with 
$S^2/Z_2$. 
In Section \ref{KKmode}, we analyze KK mode expansion of the each field in our model, and derive 
the mass spectrum of the KK modes.
Section \ref{summary} is devoted to summary and discussions. 



\section{Universal Extra dimension model on $M^4 \times S^2/Z_2$ spacetime } 
\label{S2UED}
In this section, we first recapitulate a gauge theory defined on the six-dimensional spacetime which has 
extra-space as two-sphere orbifold $S^2/Z_2$.
We then construct a six-dimensional Lagrangian for Universal Extra Dimension Model on the spacetime. 



\subsection{Gauge theory on $M^4 \times S^2/Z_2$ spacetime}
We consider a gauge theory defined on the six-dimensional spacetime $M^6$ which is assumed to be 
a direct product of the four-dimensional Minkowski spacetime $M^4$ and a compact two-sphere orbifold $S^2/Z_2$,
 such that
$M^6=M^4 \times S^2/Z_2$.
We denote the coordinate of $M^6$ by $X^M=(x^{\mu},y^{\theta}=\theta,y^{\phi}=\phi)$, 
where $x^{\mu}$ and $\{\theta,\phi \}$ are the $M^4$ coordinates and are the $S^2/Z_2$ spherical coordinates, respectively.
On the orbifold, the point $(\theta,\phi)$ is identified with $(\pi-\theta,-\phi)$.
The spacetime index $M$ runs over $\mu \in \{0,1,2,3 \}$ and $\alpha \in \{\theta,\phi\}$.
The metric of $M^6$, denoted by $g_{\scriptscriptstyle MN}$, can be written as 
\begin{equation}
g_{\scriptscriptstyle MN} = \begin{pmatrix} \eta_{\mu \nu} & 0 \\ 0 & -g_{\alpha \beta} \end{pmatrix}, 
\end{equation}
where $\eta_{\mu \nu}= diag(1,-1,-1,-1)$ and $g_{\alpha \beta}= diag(R^{2}, R^{2}\sin^{2} \theta)$ 
are metric of $M^4$ and $S^2/Z_2$ respectively, and $R$ denotes the radius of $S^2/Z_2$.
We define the vielbein $e^{M}_{A}$ 
that connects the metric of $M^6$ and that of 
the tangent space of $M^6$, denoted by $h_{AB}$, as $g_{\scriptscriptstyle MN}=e_M^{A} e_N^B h_{AB}$. 
Here $A=(\mu,a)$, where $a$ $\in$ $\{ 4,5 \}$, is the index for the coordinates of tangent space of $M^6$. 
The explicit form of the vielbeins are summarized in the Appendix.
We introduce, in this theory, a gauge field $A_{M}(x,y)=(A_{\mu}(x,y),A_{\alpha}(x,y))$, 
SO(1,5) chiral fermions $\Psi_{\pm}(x,y)$,
 and complex scalar fields $\Phi(x,y)$.
The SO(1,5) chiral fermion $\Psi_{\pm}(x,y)$ is defined by the action of SO(1,5) chiral operator $\Gamma_7$,
 which is defined as
\begin{equation}
\Gamma_7 = \gamma_5 \otimes \sigma_3,
\end{equation}
where $\gamma_5$ is SO(1,3) chiral operator and $\sigma_i(i=1,2,3)$ are Pauli matrices.
The chiral fermion $\Psi_{\pm}(x,y)$ satisfies 
\begin{equation}
\Gamma_7 \Psi_{\pm}(x,y)= \pm \Psi_{\pm}(x,y)
\end{equation}
and is obtained by acting the chiral projection operator of SO(1,5), $\Gamma_{\pm}$, on 
Dirac fermion $\Psi(x,y)$, where 
$\Gamma_{\pm}$ is defined as 
\begin{equation}
\Gamma_{\pm} = \frac{1 \pm \Gamma_7}{2}.
\end{equation}
We can also write $\Psi_{\pm}(x,y)$ in terms of SO(1,3) chiral fermion $\psi$ as 
\begin{align}
\label{chiralR}
\Psi_+ = \begin{pmatrix} \psi_R \\ \psi_L \end{pmatrix}, \\
\label{chiralL} 
\Psi_- = \begin{pmatrix} \psi_L \\ \psi_R \end{pmatrix}, 
\end{align}
where $\psi_{R(L)}$ is a right(left)-handed SO(1,3) chiral fermion. 
We should determine the boundary condition of these 
fields on $S^2/Z_2$ to specify a model.
The boundary conditions for each 
field can be defined as 
\begin{align}
\label{BC0}
\Phi(x,\pi-\theta,-\phi) &= \pm \Phi(x,\theta,\phi) \\
\label{BC1}
A_{\mu}(x,\pi-\theta,-\phi) &= A_{\mu}(x,\theta,\phi) \\
\label{BC2}
A_{\theta,\phi}(x,\pi-\theta,-\phi) &= -A_{\theta,\phi}(x,\theta,\phi) \\
\label{BC3}
\Psi(x,\pi-\theta,-\phi) &= \pm \gamma_5 \Psi(x,\theta,\phi)
\end{align}
by requiring the invariance of a six-dimensional action under the $Z_2$ transformation.

The action of the gauge theory is written, in general, as 
\begin{align}
\label{6Daction}
 S  &= \int dx^4 R^2 \sin \theta d \theta d \phi \nonumber \\ 
& \left. \times 
\biggl( \bar{\Psi}_{\pm} i \Gamma^{\mu} D_{\mu} \Psi_{\pm} + \bar{\Psi}_{\pm} i \Gamma^{a} e^{\alpha}_{a} D_{\alpha} \Psi_{\pm} 
- \frac{1}{4 g^2} g^{MN} g^{KL} Tr[F_{MK} F_{NL}] \right.  \nonumber \\
& \left. \qquad +(D^M \Phi)^* D_M \Phi -V(\Phi) -\lambda \bar{\Psi}_{\pm} \Phi \Psi_{\mp} \biggr) \right. , 
\end{align}
where $F_{MN}= \partial_M A_N(X) -\partial_N A_M(X) -[A_M(X),A_N(X)]$ is the field strength, 
$D_M$ is the covariant derivative including a spin connection, $V(\Phi)$ is the scalar potential term,
and $\Gamma_A$ represents the 6-dimensional Clifford algebra. 
Here $D_M$ and $\Gamma_A$ can be written explicitly as

\begin{align}
D_{\mu} &= \partial_{\mu} - i A_{\mu}, \\
D_{\theta} &= \partial_{\theta} - i A_{\theta}, \\
 \label{spin}
D_{\phi} &= \partial_{\phi} -i A_{\phi} \ (-i \frac{\Sigma_3}{2} \cos \theta), \\
\Gamma_{\mu} &= \gamma_{\mu} \otimes \mathbf{I}_2, \\
\Gamma_4 &= \gamma_{5} \otimes i\sigma_1, \\
\Gamma_5 &= \gamma_{5} \otimes i\sigma_2, 
\end{align}
where $ \{ \gamma_{\mu}, \gamma_{5} \} $ are the 4-dimensional Dirac matrices, 
$\mathbf{I}_d$ is $d \times d$ identity, 
and 
$\Sigma_3$ is defined as $\Sigma_3=\mathbf{I}_4 \otimes \sigma_3$.
We note that the spin connection term in $D_{\phi}$ is applied only for fermions.

We discuss the condition to obtain massless chiral fermions in four-dimensional spacetime.
The positive curvature of an extra-space gives mass to fermions in four-dimensional spacetime 
even if we introduce chiral fermions in a higher-dimensional spacetime.
The spin connection term for fermions in Eq.~(\ref{spin}) expresses the existence of positive curvature of $S^2$ and 
leads mass term of fermions in four-dimensional spacetime.
We thus need some prescription to obtain a massless fermion in four-dimensional spacetime within our model 
since $S^2$ has the positive curvature.
We then introduce a background gauge field $A^{B}_{\phi}$ which has the form 
\cite{RandjbarDaemi:1982hi,Manton:1979kb,background}
\begin{equation}
A^B_{\phi} = \hat{Q} \cos \theta
\end{equation}
where $\hat{Q}$ is a charge of some U(1) gauge symmetry, in order to cancel the mass from the curvature 
and to obtain massless fermions in four-dimensional spacetime.
Indeed, $A^B_{\phi}$ cancel the spin connection term for the upper(lower) component SO(1,3) fermion in Eq.~(\ref{chiralR})   
if the fermion has the charge $Q=+(-) \frac{1}{2}$ and the upper(lower) component gets a massless Kaluza-Klein mode.



\subsection{The Lagrangian of the model}  
Here, we specify 
our model.
Chose the gauge group $G$ as the standard model gauge group with an extra U(1)$_X$ gauge symmetry
, i.e. G=SU(3) $\times$ SU(2) $\times$ U(1)$_Y \times$ U(1)$_X$, and introduce background gauge field which 
belongs to the gauge field of the extra U(1). 
We must
 introduce the extra U(1) to obtain all the massless chiral SM fermions in $M^4$,
otherwise some SM fermions have masses, inevitably, from the positive curvature of $S^2$ when we introduce background 
gauge field which belongs to U(1)$_Y$.

We introduce fermions $Q(x,y), U(x,y), D(x,y), L(x,y)$ and $E(x,y)$ that belong to representations of 
SU(3)$\times$SU(2)$\times$U(1)$_Y$,
 which are the same as 
 the left-handed quark doublet, right-handed up-type quark, right-handed down-type quark,
left-handed lepton doublet and right-handed charged lepton.
We then assign the extra U(1) charge $Q=\frac{1}{2}$ to these fermions as the simplest case in which 
all massless SM fermions appear in four-dimensional spacetime. 
The chirality of SO(1,5) and boundary condition for these fermions are determined 
to give 
massless SM fermions in four-dimensional spacetime,
 as summarized in Table
 ~\ref{conditions}.
\begin{table}[h]
\begin{center}
\caption{SO(1,5) chirality and boundary conditions for each fermions in six dimensions.
The signs
 for boundary condition express the sign in front of $\gamma_5$ in RHS of Eq.~(\ref{BC3}).}
\label{conditions}
\begin{tabular}{|l|c|c|} \hline
Fermions & SO(1,5) chirality & boundary conditions \\ \hline
$Q(x,y)$ & $-$ & $-$ \\ \hline
$U(x,y)$ & $+$ & $+$ \\ \hline
$D(x,y)$ & $+$ & $+$ \\ \hline
$L(x,y)$ & $-$ & $-$ \\ \hline
$E(x,y)$ & $+$ & $+$ \\ \hline
\end{tabular}
\end{center}
\end{table}
The Higgs field $H(x,y)$ is introduced to 
not have U(1)$_X$ charge and to be even under the $Z_2$ action 
so that Yukawa coupling terms can be constructed.

The action of our model in six-dimensional spacetime is written as 
\begin{eqnarray}
\label{action6D}
S_{6D} = \int dx^4 R^2 \sin \theta d \theta d \phi \biggl[
(\bar{Q},\bar{U},\bar{D},\bar{L},\bar{E}) i \Gamma^M D_M (Q, U,D,L,E)^\top \nonumber \\
 -  g^{MN} g^{KL} \sum_i \frac{1}{4 g_i^2} Tr[F_{i \ MK} F_{i \ NL}] + L_{Higgs}(H) \nonumber \\  
+ [  \lambda_u Q \bar{U} H^* + \lambda_d Q \bar{D} H + \lambda_e L \bar{E} H + \textrm{h.c}]
 \biggr],
\end{eqnarray}
where $i=SU(3), SU(2), U(1)_Y$ and $U(1)_X$,
 and $L_{Higgs}(H)$ denotes a Lagrangian for Higgs field. 
The action in four-dimensional spacetime is obtained by integrating the Lagrangian over $S^2/Z_2$ coordinate.

\section{KK mode expansion and particle mass spectrum in four-dimensions}
\label{KKmode}
In this section we analyze Kaluza-Klein expansion of 
each field in our model, and derive 
mass spectrum of the Kaluza-Klein modes.



\subsection{KK mode expansion of the fermions}

The fermions $\Psi(x,y)$ can be expanded in terms of the eigenfunctions of square of Dirac operator $i\hat{D}$ on $S^2/Z_2$ where 
the Dirac operator is written as  
\begin{align}
\label{Dirac}
i \hat{D} &= i e^{\alpha a} i \sigma_a D_{\alpha} \otimes \gamma_5 \nonumber \\
&= - \frac{1}{R} \bigl[ \sigma_1 (\partial_{\theta} + \frac{\cot \theta}{2} ) 
+ \sigma_2 (\frac{1}{\sin \theta} \partial_{\phi} + i \hat{Q} \cot \theta ) \bigr] \otimes \gamma_5 , 
\end{align}
where $\hat{Q}$ is the U(1)$_X$ charge operator in our model.
We thus need to derive eigenfunctions of $(i\hat{D})^2$ first. 
The square of the Dirac operator $(i\hat{D})^2$ is written as 
\begin{align}
\label{Dirac-squire}
(-i \hat{D})^2 = \frac{1}{R^2} \bigl[ \frac{1}{\sin \theta} \partial_{\theta} (\sin \theta \partial_{\theta}) 
+ \frac{1}{\sin^2 \theta} \partial_{\phi}^2 +2i(\hat{Q} -  \frac{\sigma_3}{2})
 \frac{\cos \theta}{\sin^2 \theta} \partial_{\phi} \nonumber \\
-\frac{1}{4} -\frac{1}{4 \sin^2 \theta} +\hat{Q} \sigma_3 \frac{1}{\sin^2 \theta} - \hat{Q}^2 \cot^2 \theta \bigr].
\end{align}
We then obtain the eigenvalue equation of $(i\hat{D})^2$ as 
\begin{align}
\label{koyuu}
& \frac{1}{R^2} \bigl[ \frac{1}{\sin \theta} \partial_{\theta} (\sin \theta \partial_{\theta}) 
+ \frac{1}{\sin^2 \theta} \partial_{\phi}^2 +2i(Q -  \frac{\sigma_3}{2})
 \frac{\cos \theta}{\sin^2 \theta} \partial_{\phi} \nonumber \\
& \qquad -\frac{1}{4} -\frac{1}{4 \sin^2 \theta} + Q \sigma_3 \frac{1}{\sin^2 \theta} - Q^2 \cot^2 \theta \bigr] \Psi(\theta,\phi) 
 = -\lambda^2 \Psi(\theta,\phi),
\end{align}
where $-\lambda^2$ express the eigenvalue of $(i\hat{D})^2$ and $Q$ is the U(1)$_X$ charge of the $\Psi(\theta,\phi)$.
We expand $\Psi(\theta,\phi)$ to solve the equation such that 
\begin{equation}
\label{tenkai}
\Psi (\theta,\phi) = \sum_m \frac{ e^{i m \phi} }{\sqrt{2 \pi}}
\begin{pmatrix} \alpha_{\lambda m} (\theta) \\ \beta_{\lambda m} (\theta) \end{pmatrix} 
\end{equation}
where $m$ is an integer.
The eigenvalue equation becomes 
\begin{align}
\label{koyuu2}
& \Bigl[ \frac{d}{dz} (1-z^2) \frac{d}{dz} 
- \frac{m^2+2m(Q-\frac{\sigma_3}{2})z + (Q-\frac{\sigma_3}{2})^2}{1-z^2} \Bigr]
\begin{pmatrix} \alpha_{\lambda m} (z) \\ \beta_{\lambda m} (z) \end{pmatrix} \nonumber \\
& \left. 
\quad =-(R^2 \lambda^2  + Q^2 - \frac{1}{4}) \begin{pmatrix} \alpha_{\lambda m} (z) \\ \beta_{\lambda m} (z) \end{pmatrix}, \right.
\end{align}
where we changed the variable as $\theta \rightarrow z=\cos \theta$.
Here we note that 
the replacement of 
$m$ with $-m$ and $Q$ with $-Q$ corresponds to the exchange of 
$\alpha_{\lambda m}$ and $\beta_{\lambda m}$.
We next put $\alpha_{\lambda m}$ and $\beta_{\lambda m}$ in the following form \cite{A.A.Abrikosov},
\begin{equation}
\label{okikae}
\begin{pmatrix} \alpha_{\lambda m} (z) \\ \beta_{\lambda m} (z) \end{pmatrix}
= \begin{pmatrix} (1-z)^{\frac{1}{2}|m+Q-\frac{1}{2}|} (1+z)^{\frac{1}{2}|m-Q+\frac{1}{2}|} \xi_{\lambda m}(z) \\
(1-z)^{\frac{1}{2}|m+Q+\frac{1}{2}|} (1+z)^{\frac{1}{2}|m-Q-\frac{1}{2}|} \eta_{\lambda m}(z) \end{pmatrix}.
\end{equation}
We finally find the equation for $\eta_{\lambda m}$ and $\xi_{\lambda m}$
 by applying Eq.~(\ref{okikae}) to Eq.~(\ref{koyuu2}), as 
\begin{equation}
\label{houteishiki1}
\Bigl[ (1-z^2) \frac{d^2}{dz^2} 
-2(|m|+1)z \frac{d}{dz}-m^2-|m|+R^2 \lambda^2 \Bigr] \xi_{\lambda m}(z) = 0, 
\end{equation}
and
\begin{align}
\label{houteishiki2}
& \Bigl[ (1-z^2) \frac{d^2}{dz^2} 
+ \{ |m-1|-|m+1| -(|m-1|+|m+1|+2)z \} \frac{d}{dz}  \nonumber \\
& \left. -\frac{1}{2}m^2-\frac{1}{2}|m-1||m+1| 
-\frac{1}{2}( |m-1|+|m+1| )
-\frac{1}{2}+R^2 \lambda^2  \Bigr] \eta_{\lambda m}(z) = 0 \right. \nonumber \\
\end{align}
where we substitute $\frac{1}{2}$ 
for $Q$ since all the fermions have this charge in our model.
These equations can be attributed to the differential equation for the Jacobi polynomial $P_n^{(\alpha,\beta)}$;
the properties of the Jacobi polynomial and their differential equation are summarized in Appendix B.
We thus obtain the eigenfunctions $\xi_{\lambda m}$, $\eta_{\lambda m}$ of the form
\begin{align}
\label{xi}
\xi_{\lambda m}(z) &= C_{\xi}^{lm} P_{l-|m|}^{(|m|,|m|)}(z), \\
\label{eta}
\eta_{\lambda m}(z) &= C_{\eta}^{lm} P_{l-|m|}^{(|m+1|,|m-1|)}(z), 
\end{align} 
for $m \not= 0$, 
\begin{align}
\label{xi0}
\xi_{\lambda 0}(z) &= C_{\xi}^{l0} P_{l}^{(0,0)}(z), \\
\label{eta0}
\eta_{\lambda 0}(z) &= C_{\eta}^{l-1} P_{l-1}^{(1,1)}(z), 
\end{align} 
for $m=0$, 
and the eigenvalue $\lambda_l$ as 
\begin{equation}
\label{lambda}
\lambda = \frac{\sqrt{l(l+1)}}{R}
\end{equation}.
For any $m$, here an $l$ is integer which satisfy $l \geq m$ and $C_{\xi(\eta)}^{lm}$s are normalization constants.
The normalization of the eigenfunctions are chosen as
\begin{align}
\label{normalization1}
C_{\xi}^{lm} &= \sqrt{\frac{n! (2l+|m|+1) \Gamma(l+|m|+1)}{2^{2|m|+1} \Gamma(l+1) \Gamma(l+1)}}, \\ 
\label{normalization2}
C_{\eta}^{lm} &= i\sqrt{\frac{n! (2l-2|m|+|m+1|+|m-1|+1) \Gamma(l-|m|+|m+1|+|m-1|+1)}{2^{|m+1|+|m-1|+1}
 \Gamma(l-|m|+|m+1|+1) \Gamma(l-|m|+|m-1|+1)}}, \nonumber \\
\end{align}
so that
\begin{equation}
\int |\alpha_{lm}(z)|^2 dz = \int |\beta_{lm}(z)|^2 dz =1,  
\end{equation}
where we choose relative phase of the normalization constants as defined above for later convenience.
We therefore obtain the eigenfunctions of $(i\hat{D})^2$ as 
\begin{align}
\label{mode}
\Psi_{lm}(\theta,\phi) =  
\begin{pmatrix} \tilde{\alpha}_{lm}(z,\phi) \\
\tilde{\beta}_{lm}(z,\phi) \end{pmatrix} 
=
 \frac{e^{im\phi}}{\sqrt{2 \pi}} 
\begin{pmatrix} C_{\xi}^{lm} (1-z)^{\frac{1}{2}|m|} (1+z)^{\frac{1}{2}|m|} 
P_{l-|m|}^{(|m|,|m|)}(z)  \\
C_{\eta}^{lm} (1-z)^{\frac{1}{2}|m+1|} (1+z)^{\frac{1}{2}|m-1|} 
P_{l-|m|}^{(|m+1|,|m-1|)}(z) \end{pmatrix}, \nonumber \\
\end{align}
for $m \not= 0$ and 
\begin{equation}
\label{mode0}
\Psi_{lm}(\theta,\phi) = 
\begin{pmatrix} \tilde{\alpha}_{l0}(z) \\
\tilde{\beta}_{l0}(z) \end{pmatrix} 
 \frac{1}{\sqrt{2 \pi}} =
\begin{pmatrix} C_{\xi}^{l0} P_{l}^{(0,0)}(z)  \\
C_{\eta}^{l-10} \sqrt{1-z^2} 
P_{l-1}^{(1,1)}(z) \end{pmatrix}, 
\end{equation}
for $m=0$. 
These eigenfunctions satisfy the orthogonality relations 
\begin{equation}
\int d \Omega (\tilde{\alpha}_{lm})^* \tilde{\alpha}_{l'm'} = 
\int d \Omega (\tilde{\beta}_{lm})^* \tilde{\beta}_{l'm'} = \delta_{ll'} \delta_{mm'}.
\end{equation}
We note that the eigenfunction for $(l=0,m=0)$ has only upper component since $P_{0-1}~{(1,1)}(z)=0$.

We can obtain the KK mode functions for chiral fermions $\Psi_{\pm}(x,\theta,\phi)$
which satisfy the boundary conditions Eq.~(\ref{BC3}) in terms of 
$\tilde{\alpha}_{lm}(z,\phi)$, $\tilde{\beta}_{lm}(z,\phi)$, $\tilde{\alpha}_{l0}(z)$ and $\tilde{\beta}_{l0}(z)$ 
in Eq.~(\ref{mode}) and (\ref{mode0}).
These KK mode functions are summarized below.
\begin{enumerate}
\item The KK mode function for $\Psi_{+}(x,\theta,\phi)$ which satisfy the boundary condition 
$\Psi_{+}(x,\pi-\theta,-\phi)=\pm \gamma_5 \Psi_{+}(x,\theta,\phi)$ are
\begin{enumerate}
\item $m \not= 0$ 
\begin{align}
\label{modep}
\Psi_{+l|m|}^{(\pm \gamma_5)}(x,\theta,\phi) &= 
\begin{pmatrix} \frac{1}{\sqrt{2}} [\tilde{\alpha}_{lm}(z,\phi) \pm (-1)^{l-|m|}\tilde{\alpha}_{l-m}(z,\phi)] \psi_{R}^{l|m|}(x) \\
\frac{i}{\sqrt{2}} [\tilde{\beta}_{lm}(z,\phi) \mp (-1)^{l-|m|}\tilde{\beta}_{l-m}(z,\phi)] \psi_{L}^{l|m|}(x) \end{pmatrix} \nonumber \\
& \equiv  \begin{pmatrix} \tilde{\alpha}^{\pm}_{l|m|}(z,\phi) \psi_{R}^{l|m|}(x) \\ 
 \tilde{\beta}^{\mp}_{l|m|} (z,\phi) \psi_{L}^{l|m|}(x) \end{pmatrix},
\end{align}
\item m=0
\begin{align}
\label{modepzero}
\Psi_{+l0}^{(\pm \gamma_5)}(x,\theta,\phi) &=
\begin{pmatrix} \frac{1}{2} [\tilde{\alpha}_{l0}(z) \pm (-1)^{l}\tilde{\alpha}_{l0}(z)] \psi_{R}^{l0}(x) \\
\frac{i}{2} [\tilde{\beta}_{l-10}(z) \pm (-1)^{l}\tilde{\beta}_{l-10}(z)] \psi_{L}^{l0}(x) \end{pmatrix} \nonumber \\
 & \equiv  \begin{pmatrix} \tilde{\alpha}^{\pm}_{l0}(y,\phi) \psi_{R(L)}^{l0}(x) \\ 
 \tilde{\beta}^{\mp}_{l0} (z) \psi_{L(R)}^{l0}(x) \end{pmatrix}. 
\end{align}

\end{enumerate}

\item The KK mode function for $\Psi_{-}(x,\theta,\phi)$ which satisfy the boundary condition 
$\Psi_{-}(x,\pi-\theta,-\phi)=\pm \gamma_5 \Psi_{-}(x,\theta,\phi)$ are

\begin{enumerate}
\item $m \not= 0$ 
\begin{align}
\label{modem}
\Psi_{-l|m|}^{(\pm \gamma_5)}(x,\theta,\phi) &= 
\begin{pmatrix} \frac{1}{\sqrt{2}} [\tilde{\alpha}_{lm}(z,\phi) \mp (-1)^{l-|m|}\tilde{\alpha}_{l-m}(z,\phi)] \psi_{L}^{l|m|}(x) \\
\frac{i}{\sqrt{2}} [\tilde{\beta}_{lm}(z,\phi) \pm (-1)^{l-|m|}\tilde{\beta}_{l-m}(z,\phi)] \psi_{R}^{l|m|}(x) \end{pmatrix} \nonumber \\
&=  \begin{pmatrix} \tilde{\alpha}^{\mp}_{l|m|}(z,\phi) \psi_{L}^{l|m|}(x) \\ 
 \tilde{\beta}^{\pm}_{l|m|} (z,\phi) \psi_{R}^{l|m|}(x) \end{pmatrix}, 
\end{align}
\item m=0
\begin{align}
\label{modemzero}
\Psi_{-l0}^{(\pm \gamma_5)}(x,\theta,\phi) &=
\begin{pmatrix} \frac{1}{2} [\tilde{\alpha}_{l0}(z) \mp (-1)^{l}\tilde{\alpha}_{l0}(z)] \psi_{L}^{l0}(x) \\
\frac{i}{2} [\tilde{\beta}_{l-10}(z) \mp (-1)^{l}\tilde{\beta}_{l-10}(z)] \psi_{R}^{l0}(x) \end{pmatrix} \nonumber \\
 &= \begin{pmatrix} \tilde{\alpha}^{\mp}_{l0}(z) \psi_{L}^{l0}(x) \\ 
 \tilde{\beta}^{\pm}_{l0} (z) \psi_{R}^{l0}(x) \end{pmatrix}
\end{align}

\end{enumerate}

\end{enumerate}
where $\psi(x)$s are SO(1,3) spinors. 
We can explicitly confirm that these KK mode functions satisfy the boundary conditions by straightforward calculation using
\begin{align}
\tilde{\alpha}_{lm}(-z,-\phi) = (-1)^{l-|m|} \tilde{\alpha}_{l-m}(z,\phi), \\ 
\tilde{\beta}_{lm}(-z,-\phi) = (-1)^{l-|m|} \tilde{\beta}_{l-m}(z,\phi),
\end{align}
which are obtained by the definition of $\tilde{\alpha}_{lm}(z,\phi)$ and $\tilde{\beta}_{lm}(z,\phi)$.
We therefore expand the fermions in six-dimensional space time such that
\begin{equation}
\label{exp1}
 \Psi_+^{(\pm \gamma_5)}(x,\theta,\phi)
 = \sum_{l=0} \sum_{m=0}^l \Psi^{(\pm \gamma_5)}_{+l|m|}(x,\theta,\phi)
\end{equation}
for fermions $\Psi_{+}(x,\theta,\phi)$ which satisfy the boundary condition: 
$\Psi_{+}(x,\pi-\theta,-\phi)=\pm \gamma_5 \Psi_{+}(x,\theta,\phi)$, and 
\begin{equation}
\label{exp2}
 \Psi_-^{(\pm \gamma_5)}(x,\theta,\phi) 
 = \sum_{l=0} \sum_{m=0}^l \Psi^{(\pm \gamma_5)}_{-l|m|}(x,\theta,\phi)
\end{equation}
for fermions $\Psi_{-}(x,\theta,\phi)$ which satisfy the boundary condition: 
$\Psi_{-}(x,\pi-\theta,-\phi)=\pm \gamma_5 \Psi_{-}(x,\theta,\phi)$.
We also summarize below the action of the Dirac operator $i\hat{D}$ on the KK modes, since it is useful to analyze 
the KK mass terms.

\begin{enumerate}
\item The KK mode function for $\Psi_{+}(x,\theta,\phi)$ which satisfy the boundary condition 
$\Psi_{+}(x,\pi-\theta,-\phi)=\pm \gamma_5 \Psi_{+}(x,\theta,\phi)$ are
\begin{enumerate}
\item $m \not= 0$ 

\begin{align}
\label{sayoup}
& i\hat{D} \Psi_{+l|m|}^{(\pm \gamma_5)}(x,\theta,\phi) \nonumber \\
&= iM_l
\begin{pmatrix} -\frac{i}{\sqrt{2}} [\tilde{\alpha}_{lm}(z,\phi) \pm (-1)^{l-|m|}\tilde{\alpha}_{l-m}(z,\phi)] \psi_L^{l|m|}(x) \\
\frac{1}{\sqrt{2}} [\tilde{\beta}_{lm}(z,\phi) \mp (-1)^{l-|m|}\tilde{\beta}_{l-m}(z,\phi)] \psi_R^{l|m|}(x) \end{pmatrix}
\end{align}
\label{sayoupzero}
\item m=0
\begin{equation}
i\hat{D} \Psi_{+l0}^{(\pm \gamma_5)}(x,\theta,\phi) = iM_l
\begin{pmatrix} -\frac{i}{2} [\tilde{\alpha}_{l0}(z) \pm (-1)^{l}\tilde{\alpha}_{l0}(z)] \psi_L^{l0}(x) \\
\frac{1}{2} [\tilde{\beta}_{l-10}(z) \pm (-1)^{l}\tilde{\beta}_{l-10}(z)] \psi_R^{l0}(x) \end{pmatrix}
\end{equation}

\end{enumerate}

\item The KK mode function for $\Psi_{-}(x,\theta,\phi)$ which satisfy the boundary condition 
$\Psi_{-}(x,\pi-\theta,-\phi)=\pm \gamma_5 \Psi_{-}(x,\theta,\phi)$ are

\begin{enumerate}
\item $m \not= 0$ 
\begin{align}
\label{sayoum}
& i\hat{D} \Psi_{-l|m|}^{(\pm \gamma_5)}(x,\theta,\phi) \nonumber \\
&= iM_l
\begin{pmatrix} \frac{i}{\sqrt{2}} [\tilde{\alpha}_{lm}(z,\phi) \pm (-1)^{l-|m|}\tilde{\alpha}_{-lm}(z,\phi)] \psi_R^{l|m|}(x) \\
-\frac{1}{\sqrt{2}} [\tilde{\beta}_{lm}(z,\phi) \mp (-1)^{l-|m|}\tilde{\beta}_{l-m}(z,\phi)] \psi_L^{l|m|}(x) \end{pmatrix}
\end{align}
\item m=0
\begin{equation}
\label{sayoumzero}
i\hat{D} \Psi_{-l0}^{(\pm \gamma_5)}(x,\theta,\phi) = iM_l
\begin{pmatrix} \frac{i}{2} [\tilde{\alpha}_{l0}(z) \mp (-1)^{l}\tilde{\alpha}_{l0}(z)] \psi_R^{l0}(x) \\
\frac{-1}{2} [\tilde{\beta}_{l-10}(z) \mp (-1)^{l}\tilde{\beta}_{l-10}(z)] \psi_L^{l0}(x) \end{pmatrix}
\end{equation}

\end{enumerate}

\end{enumerate}
where $M_l = \frac{\sqrt{l(l+1)}}{R}$.
These results respect the choice of the phase between normalization constants of upper and lower components 
in Eq.~(\ref{normalization1}) and (\ref{normalization2}) 
since the Dirac operator exchange upper and lower components. 

We then derive 
the kinetic terms and 
the KK mass terms for each KK modes of the fermion in four-dimensional spacetime.
The kinetic terms for the fermion KK modes are obtained by expanding the higher-dimensional chiral fermion $\Psi_{\pm}$   
in terms of mode functions Eq.~(\ref{modep}),(\ref{modepzero}),(\ref{modem}),
 and (\ref{modemzero}) and integrating over 
$\theta$ and $\phi$.
We thus obtain the kinetic terms such that
\begin{enumerate}
\item 
For $\Psi_+^{(+\gamma_5)}(x,\theta,\phi)$ 

\begin{enumerate}
\item $m \not= 0$
\begin{equation}
\int d \Omega \bar{\Psi}^{(+\gamma_5)}_{+l|m|} i \Gamma^{\mu} \partial_{\mu} \Psi^{(+\gamma_5)}_{+l|m|}
= \bar{\psi}_R^{l|m|}(x) i \gamma^{\mu} \partial_{\mu} \psi_R^{l|m|}+\bar{\psi}_L^{l|m|}(x) i \gamma^{\mu} \partial_{\mu} \psi_L^{l|m|}
\end{equation}

\item $m = 0$
\begin{align}
& \int d \Omega \bar{\Psi}^{(+\gamma_5)}_{+00} i \Gamma^{\mu} \partial_{\mu} \Psi^{(+\gamma_5)}_{+00}
= \bar{\psi}_R^{00}(x) i \gamma^{\mu} \partial_{\mu} \psi_R^{00} \quad (l=0) \\ 
& \int d \Omega \bar{\Psi}^{(+\gamma_5)}_{+l0} i \Gamma^{\mu} \partial_{\mu} \Psi^{(+\gamma_5)}_{+l0} \nonumber \\
& \left. \ = \frac{(1+(-1)^{l})^2}{4} [\bar{\psi}_R^{l0}(x) i \gamma^{\mu} \partial_{\mu} \psi_R^{l0}+ 
\bar{\psi}_L^{l0}(x) i \gamma^{\mu} \partial_{\mu} \psi_L^{l0}] \quad (l \not= 0) \right. 
\end{align}

\end{enumerate}

\item 
For $\Psi_-^{(-\gamma_5)}(x,\theta,\phi)$ 

\begin{enumerate}
\item $m \not= 0$
\begin{equation}
\int d \Omega \bar{\Psi}^{(-\gamma_5)}_{-l|m|} i \Gamma^{\mu} \partial_{\mu} \Psi^{(-\gamma_5)}_{-l|m|}
= \bar{\psi}_R^{l|m|}(x) i \gamma^{\mu} \partial_{\mu} \psi_R^{l|m|}+\bar{\psi}_L^{l|m|}(x) i \gamma^{\mu} \partial_{\mu} \psi_L^{l|m|}
\end{equation}

\item $m = 0$
\begin{align}
& \int d \Omega \bar{\Psi}^{(-\gamma_5)}_{-00} i \Gamma^{\mu} \partial_{\mu} \Psi^{(-\gamma_5)}_{-00}
= \bar{\psi}_L^{00}(x) i \gamma^{\mu} \partial_{\mu} \psi_L^{00} \quad (l=0) \\ 
& \int d \Omega \bar{\Psi}^{(-\gamma_5)}_{-l0} i \Gamma^{\mu} \partial_{\mu} \Psi^{(-\gamma_5)}_{-l0} \nonumber \\
& = \frac{(1+(-1)^{l})^2}{4} [\bar{\psi}_R^{l0}(x) i \gamma^{\mu} \partial_{\mu} \psi_R^{l0}+ 
\bar{\psi}_L^{l0}(x) i \gamma^{\mu} \partial_{\mu} \psi_L^{l0}] \quad (l \not= 0) 
\end{align}

\end{enumerate}

\end{enumerate}
We obtain the mass terms of the fermion KK modes by using Eq.~(\ref{modep})-(\ref{modemzero}) and  Eq.~(\ref{sayoup})-(\ref{sayoumzero})
and integrating over $\theta$ and $\phi$, such that
\begin{enumerate}
\item 
For $\Psi_+^{(+\gamma_5)}(x,\theta,\phi)$ 

\begin{enumerate}
\item $m \not= 0$
\begin{equation}
\int d \Omega \bar{\Psi}^{(+\gamma_5)}_{+l|m|} i \hat{D} \Psi^{(+\gamma_5)}_{+l|m|}
= M_l[\bar{\psi}_R^{l|m|}(x) \psi_L^{l|m|}+\bar{\psi}_L^{l|m|}(x) \psi_R^{l|m|}]
\end{equation}

\item $m = 0$
\begin{align}
& \int d \Omega \bar{\Psi}^{(+\gamma_5)}_{+00} i \hat{D} \Psi^{(+\gamma_5)}_{+00}
= 0 \quad ( l=0) \\ 
& \int d \Omega \bar{\Psi}^{(+\gamma_5)}_{+l0} i \hat{D} \Psi^{(+\gamma_5)}_{+l0} \nonumber \\
& = \frac{(1+(-1)^{l})^2}{4} M_l [\bar{\psi}_R^{l0}(x) \psi_L^{l0}+ 
\bar{\psi}_L^{l0}(x) \psi_R^{l0}] \quad ( l \not= 0) 
\end{align}

\end{enumerate}

\item 
For $\Psi_-^{(-\gamma_5)}(x,\theta,\phi)$ 

\begin{enumerate}
\item $m \not= 0$
\begin{equation}
\int d \Omega \bar{\Psi}^{(-\gamma_5)}_{-l|m|} i \hat{D} \Psi^{(-\gamma_5)}_{-l|m|}
= -M_l [\bar{\psi}_R^{l|m|}(x) \psi_L^{l|m|}+\bar{\psi}_L^{l|m|}(x) \psi_R^{l|m|}]
\end{equation}

\item $m = 0$
\begin{align}
& \int d \Omega \bar{\Psi}^{(-\gamma_5)}_{-00} i \hat{D} \Psi^{(-\gamma_5)}_{-00}
= 0 \quad ( l=0) \\ 
& \int d \Omega \bar{\Psi}^{(-\gamma_5)}_{-l0} i \hat{D} \Psi^{(-\gamma_5)}_{-l0} \nonumber \\
&= -\frac{(1+(-1)^{l})^2}{4} M_l [\bar{\psi}_R^{l0}(x) \psi_L^{l0}+ 
\bar{\psi}_L^{l0}(x) \psi_R^{l0}] \quad ( l \not= 0). 
\end{align}

\end{enumerate}

\end{enumerate}
We have thus confirmed that the fermion $\Psi_-^{(-\gamma_5)}(x,\theta,\phi)$($\Psi_+^{(+\gamma_5)}(x,\theta,\phi)$) has 
the chiral left(right)-handed massless mode. 

We must consider Yukawa coupling of Higgs zero mode and fermion KK modes to obtain mass spectrum of the KK particles after the 
electroweak symmetry breaking. 
The Yukawa coupling term in six-dimensional spacetime has the form
 
\begin{equation}
L_{Yukawa} = \int d \Omega \bigl[ \lambda \bar{\Psi}^{(+\gamma_5)}_+(x,\theta,\phi)
  H (x,\theta,\phi) \Psi_-^{(-\gamma_5)} (x,\theta,\phi) + \textrm{h.c} \bigr] ,
\end{equation}
and we obtain the coupling of Higgs zero mode and fermion KK mode in four-dimensional spacetime as
\begin{equation}
L_{Yukawa0} 
 = \sum_{l m} \lambda \biggl[ 
 \bar{\psi}_R^{l|m|}(x) H^{00}(x) \tilde{\psi}_L^{l|m|}(x) +
 \bar{\psi}_L^{l|m|}(x) H^{00}(x) \tilde{\psi}_R^{l|m|}(x) \biggr] +\textrm{h.c}, 
\end{equation}
where we put tilde on fermions which are obtained from $\Psi_-^{(-\gamma_5)}$.
After the electroweak symmetry breaking, the Higgs zero mode have a vacuum expectation value(v.e.v) and 
we have the mass term of the kk mode of the form 
\begin{equation}
\begin{pmatrix} \bar{\psi}_{lm} & \bar{\tilde{\psi}}_{lm} \end{pmatrix} 
\begin{pmatrix} M_l & m_f \\ m_f & -M_l \end{pmatrix} 
\begin{pmatrix} \psi_{lm} \\ \tilde{\psi}_{lm} \end{pmatrix}
\end{equation}
where $m_f$s express the masses of the SM fermions originated from the Yukawa coupling term.
Since this mass term mix $\psi$ and $\tilde{\psi}$ we must diagonalize the mass term.
We change the basis of $\psi$ and $\tilde{\psi}$ as 
\begin{equation}
\begin{pmatrix} \psi_{lm} \\ \tilde{\psi}_{lm} \end{pmatrix} = 
\begin{pmatrix} \gamma_5 \cos \alpha_l & \sin \alpha_l \\ -\gamma_5 \sin \alpha_l  & \cos \alpha_l \end{pmatrix} 
\begin{pmatrix} \psi'_{lm} \\ \tilde{\psi}'_{lm} \end{pmatrix}
\end{equation}
to diagonalize the mass term, where 
\begin{equation}
\tan 2 \alpha_l = \frac{m_f}{M_l}.
\end{equation}
After diagonalizing mass term, we obtain the mass spectrum 
\begin{equation}
\label{massf}
M_f^{l} = \pm \sqrt{M_l^2 + m_f^2}.
\end{equation}
We note that the KK mass $M_l$ do not depend on $m$ and $m$s are not mixed in mass terms, so that 
degeneracy of KK masses is 
\begin{align}
\label{degeneracy1}
& l+1 \qquad \textrm{for} \qquad l=\textrm{even}, \\
\label{degeneracy2}
& l \qquad \textrm{for} \qquad l=\textrm{odd},
\end{align}
since $m$ runs $0$ to $l$.



\subsection{KK mode expansion of gauge field}

Let us focus on the quadratic terms of gauge field Lagrangian in (\ref{6Daction}) 
since we would like to know the mass spectrum of gauge fields. 
Decomposing the Lagrangian into 4D components, we obtain
\begin{align}
& L_{{\rm gauge}}^{{\rm quadratic}} \nonumber \\
& =
-\frac{1}{4g^2} \sin \theta 
\left[
R^2 (\partial_\mu A_\nu - \partial_\nu A_\mu)(\partial^\mu A^\nu - \partial^\nu A^\mu) \right. \nonumber \\
& \left. 
-2 \left\{
(\partial_\mu A_\theta)(\partial^\mu A_\theta) -2 (\partial_\mu A_\theta)(\partial_\theta A^\mu) 
+ (\partial_\theta A_\mu)(\partial_\theta A^\mu)
\right\} \right. \nonumber \\
& \left. 
-2 \left\{
(\partial_\mu \tilde{A}_\phi)(\partial^\mu \tilde{A}_\phi) - \frac{2}{\sin \theta} (\partial_\mu \tilde{A}_\phi)(\partial_\phi A^\mu) 
+ \frac{1}{\sin^2 \theta}(\partial_\phi A_\mu)(\partial_\phi A^\mu)
\right\} \right. \nonumber \\
& \left. 
+ \frac{2}{R^2 \sin^2 \theta} \left\{
(\partial_\theta \sin \theta \tilde{A}_\phi)(\partial_\theta \sin \theta \tilde{A}_\phi)
 -2 (\partial_\theta \sin \theta \tilde{A}_\phi)(\partial_\phi A_\theta) 
+ (\partial_\phi A_\theta)(\partial_\phi A_\theta)
\right\}
\right] \nonumber \\ 
\label{gauge}
\end{align}
where we defined $\tilde{A}_\phi$ as $\tilde{A}_\phi \equiv A_\phi/\sin \theta$ for the kinetic term to be canonical. 
We note that the background field $\langle A_\phi \rangle$ belongs to the U(1)$_X$ gauge field 
and hence $[A_{\mu, \theta}, \langle A_\phi \rangle] = 0$. 
Namely, we have no background gauge field which induces masses of $A_\mu$ and $A_\theta$. 

In order to fix the gauge, the follwing gauge-fixing Lagrangian 
that 
cancels the mixing terms $A_\mu$ and $A_\theta, \tilde{A}_\phi$is added. 
\begin{align}
L_{{\rm gf}} &= - \sqrt{-g} \frac{1}{2\xi g^2} 
\left[
\partial_\mu A^\mu + \frac{\xi}{R^2 \sin \theta} \left(\partial_\theta (\sin \theta A^\theta) 
+ \frac{1}{\sin \theta} \partial_\phi A^\phi \right)
\right]^2 \nonumber \\
&= - \frac{R^2 \sin \theta}{2\xi g^2} 
\left[
(\partial_\mu A^\mu)^2 + \frac{\xi^2}{R^4 \sin^2\theta}
\left(\partial_\theta(\sin \theta A_\theta) 
+ \frac{1}{\sin \theta} \partial_\phi A_\phi 
\right)^2 \right. \nonumber \\
& \qquad \left. 
- \frac{2\xi}{R^2\sin \theta} (\partial_\mu A^\mu) 
\left(
\partial_\theta (\sin \theta A_\theta) + \frac{1}{\sin \theta} \partial_\phi A_\phi
\right)
\right]
\label{gf}
\end{align}
where $\xi$ is a gauge-fixing parameter. 

Combining (\ref{gauge}) and (\ref{gf}) and after partial integration, we obtain 
\begin{align}
& L_{{\rm gauge}}^{{\rm quadratic}} + L_{{\rm gf}} \nonumber \\ 
& =
-\frac{1}{4g^2} \sin \theta 
\left[
R^2 (\partial_\mu A_\nu - \partial_\nu A_\mu)(\partial^\mu A^\nu - \partial^\nu A^\mu) \right. \nonumber \\
& \left. -2 \left\{
(\partial_\mu A_\theta)(\partial^\mu A_\theta) 
+ (\partial_\theta A_\mu)(\partial_\theta A^\mu)
+ (\partial_\mu \tilde{A}_\phi)(\partial^\mu \tilde{A}_\phi) 
+ \frac{1}{\sin^2 \theta}(\partial_\phi A_\mu)(\partial_\phi A^\mu)
\right\} \right. \nonumber \\
& \left. 
+ \frac{2}{R^2 \sin^2 \theta} \left\{
(\partial_\theta \sin \theta \tilde{A}_\phi)(\partial_\theta \sin \theta \tilde{A}_\phi) 
-2 (\partial_\theta \sin \theta \tilde{A}_\phi)(\partial_\phi A_\theta) 
+ (\partial_\phi A_\theta)(\partial_\phi A_\theta)
\right\}
\right] \nonumber \\
&- \frac{R^2 \sin \theta}{2\xi g^2} 
\left[
(\partial_\mu A^\mu)^2 + \frac{\xi^2}{R^4 \sin^2\theta}
\left(\partial_\theta(\sin \theta A_\theta) 
+ \partial_\phi \tilde{A}_\phi 
\right)^2 
\right] \nonumber \\
& = 
-\frac{1}{4g^2} \sin \theta 
\left[
R^2 (\partial_\mu A_\nu - \partial_\nu A_\mu)(\partial^\mu A^\nu - \partial^\nu A^\mu) \right. \nonumber \\ 
& \left. -2 \left\{
 (\partial_\theta A_\mu)(\partial_\theta A^\mu) 
+ \frac{1}{\sin^2 \theta}(\partial_\phi A_\mu)(\partial_\phi A^\mu)
\right\}
\right. \nonumber \\
& \left. -2 \left\{
(\partial_\mu A_\theta)(\partial^\mu A_\theta) 
+ (\partial_\mu \tilde{A}_\phi)(\partial^\mu \tilde{A}_\phi) 
\right\}  
+ \frac{2}{R^2 \sin^2 \theta} \left(
(\partial_\theta \sin \theta \tilde{A}_\phi)-(\partial_\phi A_\theta )
\right)^2
\right]  \nonumber \\
&- \frac{R^2 \sin \theta}{2\xi g^2} 
\left[
(\partial_\mu A^\mu)^2 + \frac{\xi^2}{R^4 \sin^2\theta}
\left(\partial_\theta(\sin \theta A_\theta) 
+ \partial_\phi \tilde{A}_\phi 
\right)^2 
\right]. 
\label{gaugegf}
\end{align}
We find that KK mass term for the four-dimensional components of gauge field can be diagonalized by 
expanding them by spherical harmonics since the extra-kinetic terms can be expressed by the 
square of angular momentum operator.
Extra-components of guage field, however, do not have clear form of extra-kinetic terms to be diagonalized.
We then perform following substitutions, 
\begin{eqnarray}
\label{okikae1}
\tilde{A}_{\phi}(x,\theta,\phi) &=& \partial_{\theta} \phi_1(x,\theta,\phi) 
+ \frac{1}{\sin \theta} \partial_{\phi} \phi_2(x,\theta,\phi) \\
\label{okikae2}
A_{\theta}(x,\theta,\phi) &=& \partial_{\theta} \phi_2(x,\theta,\phi) 
- \frac{1}{\sin \theta} \partial_{\phi} \phi_1(x,\theta,\phi).
\end{eqnarray}
Here we note that there is no component which is independent of $S^2/Z_2$ coordinates since 
they are forbidden by the boundary conditions.
This substitution leads 

\begin{eqnarray}
\label{henkei1}
 \frac{1}{\sin \theta} (\partial_{\theta} \sin \theta \tilde{A}_{\phi}) 
-\frac{1}{\sin \theta} \partial_{\phi} A_{\theta} 
= \frac{1}{\sin \theta} \partial_{\theta}(\sin \theta \partial_{\theta} \phi_1) + \frac{1}{\sin^2 \theta} \partial_{\phi}^2 \phi_1, \\
\label{henkei2}
 \frac{1}{\sin \theta} (\partial_{\theta} \sin \theta A_{\theta}) 
+\frac{1}{\sin \theta} \partial_{\phi} \tilde{A}_{\phi} 
= \frac{1}{\sin \theta} \partial_{\theta}(\sin \theta \partial_{\theta} \phi_2) + \frac{1}{\sin^2 \theta} \partial_{\phi}^2 \phi_2,
\end{eqnarray}
where RHS's are expressed by square of angular momentum opperator acting on $\phi_{1(2)}$.
The four-dimensional kinetic term of $A_{\phi}$ and $A_{\theta}$ is also rewritten as

\begin{eqnarray}
& & \sin \theta [\partial_{\mu} A_{\theta} \partial^{\mu} A_{\theta}+ \partial_{\mu} \tilde{A}_{\phi} \partial^{\mu} \tilde{A}_{\phi}] 
\nonumber \\
&& \left. = \sin \theta \biggl[ -\partial_{\mu} \phi_1 \partial^{\mu} \Bigl[
\frac{1}{\sin \theta} \partial_{\theta}(\sin \theta \partial_{\theta} \phi_1) + \frac{1}{\sin^2 \theta} \partial_{\phi}^2 \phi_1 \Bigr]
\right.
\nonumber \\ 
&& \left. \qquad \qquad
-\partial_{\mu} \phi_2 \partial^{\mu} \Bigl[
\frac{1}{\sin \theta} \partial_{\theta}(\sin \theta \partial_{\theta} \phi_2) + \frac{1}{\sin^2 \theta} \partial_{\phi}^2 \phi_2 \Bigr]
\biggr] \right. \nonumber \\
&& \left. \qquad \qquad -2 \partial_{\mu}(\partial_{\theta} \phi_2) \partial^{\mu}(\partial_{\phi} \phi_1) 
+2 \partial_{\mu}(\partial_{\theta} \phi_1) \partial^{\mu}(\partial_{\phi} \phi_2). \right. 
\end{eqnarray}

Note that the last two terms are canceled between them after expanding $\phi_{1(2)}$ by spherical harmonics 
and performing partial integration.
After substitution Eq.~(\ref{okikae1}) and (\ref{okikae2}),   
we obtain 
\begin{align}
& L_{{\rm gauge}}^{{\rm quadratic}} + L_{{\rm gf}} \nonumber \\ 
& = 
-\frac{1}{4g^2} \sin \theta 
\left[
R^2 (\partial_\mu A_\nu - \partial_\nu A_\mu)(\partial^\mu A^\nu - \partial^\nu A^\mu)  - 2 A^{\mu} \hat{L}^2 A_{\mu} 
\right. \nonumber \\
& \left. +2 \left\{
\partial_\mu \phi_1\partial^\mu(\hat{L}^2 \phi_1)+ \partial_\mu \phi_2\partial^\mu(\hat{L}^2 \phi_2)
\right\}  
- \frac{2}{R^2}  (\hat{L}^2 \phi_1)^2
\right]  \nonumber \\
&- \frac{R^2 \sin \theta}{2\xi g^2} 
\left[
(\partial_\mu A^\mu)^2 + \frac{\xi^2}{R^4 } (\hat{L}^2 \phi_2)^2
\right],
\label{gaugegf2}
\end{align}

where $\hat{L}^2=-(1/\sin \theta) \partial_{\theta}(\sin \theta \partial \theta)-(1/\sin^2) \theta \partial_{\phi}^2$ is the 
square of angular momentum operator.
It is now clear that diagonal KK mass terms can be obtained by expanding gauge fields using spherical harmonics.
The mode expansions are then carried out in the following way. 
\begin{align}
\label{exp3}
A_\mu(x,\theta,\phi) &= \sum_{l,m} A_\mu^{lm}(x) Y_{lm}^+(\theta,\phi), \\
\label{exp4}
\phi_{1(2)}(x, \theta, \phi) &=   
\sum_{l(\not=0),m} \phi_{1(2)}^{lm}(x) \frac{Y_{lm}^+(\theta, \phi)}{\sqrt{l(l+1)}} 
\end{align}
where the mode finction $Y_{lm}^\pm(\theta, \phi)$ is defined as 
\begin{align}
\label{modef1}
Y_{lm}^+(\theta, \phi) & \equiv  \frac{(i)^{l+m}}{\sqrt{2}}[Y_{lm}(\theta, \phi) \pm (-1)^{l} Y_{l-m}(\theta, \phi)] 
\quad \textrm{for} \quad m \not=0  \\
\label{modef2}
Y_{lm}^-(\theta, \phi) & \equiv  \frac{(i)^{l+m+1}}{\sqrt{2}}[Y_{lm}(\theta, \phi) \pm (-1)^{l} Y_{l-m}(\theta, \phi)] 
\quad \textrm{for} \quad m \not=0  \\
\label{modef3}
Y_{l0}^{+(-)}(\theta) & \equiv Y_{l0}(\theta) \quad \textrm{for} \quad m =0 \ \textrm{and} \ l=\textrm{even(odd)} \nonumber \\
& \equiv 0 \qquad \quad \textrm{for} \quad m=0 \ \textrm{and} \ l=\textrm{odd(even)}
\end{align}
We note that the mode functions $Y_{lm}^\pm$ are eigenfunctions with $Z_2$ parity $\pm$ 
under $Z_2$ action $(\theta, \phi) \to (\pi-\theta, -\phi)$ 
because of the property $Y_{lm}(\pi-\theta, -\phi) = (-1)^l Y_{l-m}(\theta, \phi)$. 
We further notice that 
the phase factors $(i)^{l+m(+1)}$ in the mode functions Eq.~(\ref{modef1}) and (\ref{modef2}) are required 
since the fields $A_{\mu}$ and $\phi_{1(2)}$ must be real and
the spherical harmonics satisfy $(Y_{lm})^*=(-1)^{m}Y_{l-m}$.

Substituting the mode expansions into the Lagrangian (\ref{gaugegf2}) 
and integrating out $\theta, \phi$ coordinates leads to
\begin{align}
& {L}_{{\rm gauge}}^{{\rm quadratic}} + {L}_{{\rm gf}} \nonumber \\
& =
-\frac{1}{4} \sum_{l,m(\ne 0)} (\partial_\mu A_\nu^{lm} - \partial_\nu A_\mu^{lm})
(\partial^\mu A^{\nu lm} - \partial^\nu A^{\mu lm}) \nonumber \\
&- \frac{1}{4} \sum_{l:{\rm even}} (\partial_\mu A_\nu^{l0} - \partial_\nu A_\mu^{l0})
(\partial^\mu A^{\nu l0} - \partial^\nu A^{\mu l0}) \nonumber \\
&+ \frac{1}{2} \sum_{l,m(\ne0)} (\partial_\mu \phi_1^{lm})(\partial^\mu \phi_1^{lm}) 
+ \sum_{l(\not=0):{\rm even}} (\partial_\mu \phi_1^{l0})(\partial^\mu \phi_1^{l0}) \nonumber \\
&
+ \frac{1}{2} \sum_{l,m(\ne0)} (\partial_\mu \phi_2^{lm})(\partial^\mu \phi_2^{lm}) 
+ \sum_{l(\not=0):{\rm even}} (\partial_\mu \phi_2^{l0})(\partial^\mu \phi_2^{l0})
\nonumber \\  
&
+ \sum_{l,m} \frac{l(l+1)}{2R^2} A_\mu^{lm} A^{\mu lm} 
-  \sum_{l,m(\ne0)} \frac{l(l+1)}{2R^2}(\phi_1^{lm})^2
- \xi \sum_{l,m(\ne0)} \frac{l(l+1)}{2R^2} (\phi_2^{lm})^2 
\nonumber \\
&- \sum_{l(\not=0):{\rm even}} \frac{l(l+1)}{2R^2} 
\left[ (\phi_1^{l0})^2 + \xi (\phi_2^{l0})^2 \right] 
- \frac{1}{2\xi} \sum_{lm(\ne 0)} 
(\partial_\mu A^{\mu lm})^2 
- \frac{1}{2\xi} \sum_{l:\textrm{even}} (\partial_\mu A^{\mu l0})^2  \nonumber \\ 
\label{4Dqdrtc}
\end{align}
A rescaling $A_\mu \to R^{-1} A_\mu$ was done 
so that the gauge kinetic term 
is canonical. 
The KK modes of the $\phi_2$ are interpreted as Nambu-Goldstone(NG) bosons since their KK masses are proportional to 
the gauge fixing parameter $\xi$.
These NG bosons will be eaten by KK modes of four dimensional components of gauge field giving their longitudinal component.

Next, let us turn to the Higgs part to 
incorporate the electroweak symmetry breaking effects.  
Higgs part of 
the Lagrangian is given by 
\begin{align}
 L_{{\rm Higgs}} &= \sqrt{-g} \left[ g^{MN} (D_M H)^\dag D_N H - V(H) \right] \nonumber \\
&= R^2 \sin \theta
\left[
\eta^{\mu \nu} (D_\mu H)^\dag D_\nu H \right. \nonumber \\
& \left. - \frac{1}{R^2} |D_\theta H|^2 - \frac{1}{R^2 \sin^2 \theta} |D_\phi H|^2 
-V(H) \right], \\
D_M &= \partial_M -ig_2 A_M -\frac{i}{2} g_1 B_M  
\end{align}
where $g_{1,2}$ and $A_M, B_M$ are the gauge coupling constants and gauge fields 
of $SU(2)_L, U(1)_Y$ gauge groups, respectively. 
$V(H)$ denotes a Higgs potential. 
The gauge boson masses are obtained from the covariant derivatives as in the standard model 
by putting the Higgs VEV $H^T=(0, \frac{v}{\sqrt{2}})$, 
\begin{align}
L_{{\rm Higgs}} &\supset R^2 \sin \theta \frac{1}{4} \frac{1}{2} \left|
\left(
\begin{array}{cc}
g_2 A_M^3+ g_1 B_M & g_2(A_M^1 -i A_M^2) \\
g_2(A_M^1 +i A_M^2) & -g_2 A_M^3 + g_1 B_M \\
\end{array}
\right)
\left(
\begin{array}{c}
0\\
v\\
\end{array}
\right) \right|^2 \nonumber \\
&= R^2 \sin \theta \left[ m_W^2 W_\mu^+ W^{\mu-} + \frac{1}{2} m_Z^2 Z_\mu Z^\mu \right. \nonumber \\ 
& \left. \qquad \qquad -\frac{1}{R^2} \left(
m_W^2 W_\theta^+ W_\theta^{-} + \frac{1}{2} m_Z^2 Z_\theta Z_\theta
\right) \right. \nonumber \\
& \left. \qquad \qquad -\frac{1}{R^2} \left(
m_W^2 \tilde{W}_\phi^+ \tilde{W}_\phi^{-} + \frac{1}{2} m_Z^2 \tilde{Z}_\phi \tilde{Z}_\phi
\right)
\right] \nonumber \\
&= \sum_{l,m} \left[ m_W^2 W_\mu^{+lm} W^{\mu- lm} + \frac{1}{2} m_Z^2 Z_\mu^{lm} Z^{\mu lm} \right. \nonumber \\
& \left. \qquad \qquad - \left(
m_W^2 W_1^{+lm} W_1^{-lm} + \frac{1}{2} m_Z^2 Z_1^{lm} Z_1^{lm}
\right) \right. \nonumber \\
& \left. \qquad \qquad - \left(
m_W^2 W_2^{+lm} W_2^{-lm} + \frac{1}{2} m_Z^2 Z_2^{lm} Z_2^{lm}
\right)
\right]
\label{WZ}
\end{align}
where $W_{1(2)}$ and $Z_{1(2)}$ are defined by the substitution Eq.~(\ref{okikae1}) and (\ref{okikae2}).

Combining the results (\ref{4Dqdrtc}) and (\ref{WZ}),
the mass spectrum of SU(2)$_L$ $\times$ U(1)$_Y$ gauge sector is summarized as follows. 
\begin{align}
\label{massg}
&W_\mu^{lm}: m_W^2 + \frac{l(l+1)}{R^2}, \quad Z_\mu^{lm}: m_Z^2 + \frac{l(l+1)}{R^2}, \quad \gamma_\mu^{lm}:  \frac{l(l+1)}{R^2}, 
\nonumber \\
&W_{1}^{lm}: m_W^2 + \frac{l(l+1)}{R^2}, \quad Z_{1}^{lm}: m_Z^2 + \frac{l(l+1)}{R^2}, 
\quad \gamma_{1}^{lm}:  \frac{l(l+1)}{R^2},  \nonumber \\
&W_{2}^{lm}: m_W^2 + \xi \frac{l(l+1)}{R^2}, \quad Z_{2}^{lm}: m_Z^2 + \xi \frac{l(l+1)}{R^2}, 
\quad \gamma_{2}^{lm}: \xi \frac{l(l+1)}{R^2}.  
\end{align}

Here we mention extra U(1)$_X$ sector in our model.
We notice that the U(1)$_X$ symmetry is anomalous and is broken at the quantum level, so that 
its gauge boson should be heavy \cite{Scrucca:2003ra}. 
We thus expect the U(1)$_X$ gauge boson and its KK modes are decoupled from the low energy sector of our model.

We, therefore, conclude that the lightest KK particles are 1st KK mode of four-dimensional components of massless gauge bosons
and that of physical scalar boson originated from extra components of gauge field. 
These kk particles are the 1st KK mode of photon $\gamma_{\mu}^{11}$, scalar photon $\gamma_1^{11}$, 
gluon $g_{\mu}^{11}$ and scalar gluon $g_1^{11}$ at tree level. 
We can also guess that the 1st KK photon is the promising candidate of the lightest KK particle after including a 
quantum correction and it would be a good candidate for the dark matter \cite{Cheng:2002iz}.




\subsection{KK mode expansion of the Higgs field}
Here we discuss 
the KK mode expansion and mass spectrum of the Higgs field.
We thus focus on the kinetic-mass terms of the Higgs field in six-dimensional space time.
The kinetic-mass term has the form
\begin{align}
 L_{Higgs-kinetic}^{(6D)} 
&= R^2 \sin \theta \bigl[\partial^{\mu} H^{\dagger}(X) \partial_{\mu} H(X) \nonumber \\
& -\frac{1}{R^2} \partial_{\theta} H^{\dagger}(X) \partial_{\theta} H(X) 
-\frac{1}{R^2 \sin^2 \theta} \partial_{\phi} H^{\dagger}(X) \partial_{\phi} H(X)   \bigr].
\end{align}
After partial integration, we obtain
\begin{align} 
\label{Lhiggs}
 L_{Higgs-kinetic}^{(6D)} &= R^2 \sin \theta \biggl[\partial^{\mu} H^{\dagger}(X) \partial_{\mu} H(X) \nonumber \\
& +\frac{1}{R^2} H^{\dagger}(X) \biggl( \frac{1}{\sin \theta} \partial_{\theta}(\sin \theta \partial_{\theta})  
+\frac{1}{\sin^2 \theta} \partial_{\phi}^2 \biggr) H(X)   \biggr],
\end{align}
where the derivative operator in 
the second term has the form of square of angular momentum operator. 

The mode expansion of the Higgs field is carried out as follows,
\begin{equation}
\label{exp6}
H(x,\theta,\phi) = \sum_{l,m} H^{lm}(x) Y_{lm}^+(\theta,\phi)
\end{equation}
since the Higgs field has even parity under the $Z_2$ action.

Substituting the mode expansion into the Lagrangian Eq.~(\ref{Lhiggs}) and integrating out $\theta 
and \phi$ coordinates, we find
the kinetic and mass terms of the Higgs field in four-dimensional spacetime

\begin{equation}
L_{Higgs-kinetic-mass}^{(4D)} = \partial^{\mu} (H^{lm}(x))^{\dagger} \partial_{\mu} H^{lm}(x) 
- \frac{l(l+1)}{R^2} (H^{lm}(x))^{\dagger} H^{lm}(x).
\end{equation}
We, therefore, find the mass spectrum of the Higgs field such that 
\begin{equation}
\label{massh}
M_l = \sqrt{\frac{l(l+1)}{R^2}+m_H^2},
\end{equation}
where $m_H$ is the Higgs zero mode mass obtained from the Higgs potential after electro weak symmetry breaking.
There are $l+1(l)$ mass degeneracies of the KK modes for even(odd) $l$ since $m$ runs from $0$ to $l$ for each $l$
and $Y^+_{l0}=0$ for $l=$odd.



\subsection{The KK-parity for each KK modes}
We discuss the KK-parity for each KK modes to investigate the stability of the lightest KK particle.
In our model, the KK momentum is not conserved due to the orbifolding, 
but the discrete part is still conserved as a remnant of KK momentum conservation. 
We can see that there is an additional discrete $Z_2'$ symmetry of $(\theta,\phi) \rightarrow (\theta,\phi + \pi)$,
which is different from the previous $Z_2$ symmetry. 
This $Z_2'$ symmetry is understood as 
the symmetry of interactions under the exchange of two fixed points on $S^2/Z_2$ orbifold which are
 points 
$(\frac{\pi}{2},0)$ and $(\frac{\pi}{2},\pi)$.
Note that the fixed points have different $\phi$ coordinates $0$ and $\pi$ but 
the same $\theta$ coordinate $\pi/2$ 
so that the $Z_2'$ action shift only 
in the $\phi$ coordinate.
The mode functions for each fields in Eqs.~(\ref{exp1}),(\ref{exp2}),(\ref{exp3}),(\ref{exp4}),
and (\ref{exp6}) are transformed under the $Z_2'$ action $(\theta,\phi) \rightarrow (\theta,\phi + \pi)$ such that  
\begin{align}
\Psi_{+l|m|}^{\pm \gamma_5}(x,\theta,\phi+\pi) &= (-1)^{m} \Psi_{+l|m|}^{\pm \gamma_5}(x,\theta,\phi), \\
\Psi_{-l|m|}^{\pm \gamma_5}(x,\theta,\phi+\pi) &= (-1)^{m} \Psi_{+l|m|}^{\pm \gamma_5}(x,\theta,\phi), \\
Y_{lm}^{\pm}(\theta,\phi+\pi) &= (-1)^m Y_{lm}^{\pm}(\theta,\phi+\pi).
\end{align}
Thus the KK-parity is defined as $(-1)^m$ and we find the KK-parity is conserved as a consequence of the $Z_2'$ symmetry of 
the Lagrangian in six-dimensional spacetime.
We, therefore, can confirm the stability of the lightest KK particle which would be the 1st KK mode of photon $\gamma_{\mu}^{11}$  
since this mode has $m=1$ and cannot decay into SM particles.
We summarize the KK particle masses, mass degeneracy and the KK parity in Table.~\ref{mass-spectrum}

\begin{table}[h]
\begin{center}
\caption{Summary of the KK particle masses, mass degeneracy and the KK parity for KK modes of fermion $\psi^{lm}$, 
four dimensional gauge bosons $A^{lm}_{\mu}$, scalar gauge bosons $\phi^{lm}_{1(2)}$
 and Higgs boson $H^{lm}$, where the $m_f$, $m_g$ and $m_H$ are the zero mode masses 
for the fermions, the gauge bosons and the Higgs boson respectively, which correspond to the masses of SM particles.
\label{mass-spectrum}}
\renewcommand{\arraystretch}{2}
\begin{tabular}{|l|c|l|c|} \hline
particle  & KK mass$^2$ & mass degeneracy & KK parity \\ \hline
$\psi^{lm}(x)$             & $\frac{l(l+1)}{R^2}+m_f^2$                & $l+1$ for $l=$even ($0 \leq m \leq l$)  
& $(-1)^m$ \\ 
                           &                                                  & $l$ for $l=$odd ($0 < m \leq l$)        
& \\ \hline
$A^{lm}_{\mu}(x)$          & $\frac{l(l+1)}{R^2}+m_g^2$                & $l+1$ for $l=$even ($0 \leq m \leq l$)  
& $(-1)^m$ \\ 
                           &                                           & $l$ for $l=$odd ($0 < m \leq l$)     
& \\ \hline 
$\phi_1^{lm}(x)$           & $\frac{l(l+1)}{R^2}+m_g^2$                & $l+1$ for $l=$even($\not=0$) ($0 \leq m \leq l$)  
& $(-1)^m$ \\ 
                           &                                           & $l$ for $l=$odd ($0 < m \leq l$)
& \\ \hline
$\phi_2^{lm}(x)$           & $\xi \frac{l(l+1)}{R^2}+m_g^2$          & $l+1$ for $l=$even($\not=0$) ($0 \leq m \leq l$)  
& $(-1)^m$ \\ 
                           &                                           & $l$ for $l=$odd ($0 < m \leq l$)
& \\ \hline
$H^{lm}(x)$                & $\frac{l(l+1)}{R^2}+m_H^2$                & $l+1$ for $l=$even ($0 \leq m \leq l$)            
& $(-1)^m$ \\
                           &                                                  & $l$ for $l=$odd ($0 < m \leq l$)               
& \\ \hline          
\end{tabular}
\end{center}
\end{table}



\section{Summary and discussions}
\label{summary}

We proposed a new universal extra dimension model which is defined on 
the six-dimensional spacetime whose extra space is the two-sphere orbifold $S^2/Z_2$ and analyzed 
the mass spectrum of Kaluza-Klein particles in the model.

We first specified our model in six-dimensional spacetime $M^4 \times S^2/Z_2$.
The orbifold $S^2/Z_2$ is clarified by operating the $Z_2$ action on $S^2$ and 
the feature of the gauge theory on $M^4 \times S^2/Z_2$ is summarized in section 2.
There,
we mentioned that a massless mode of fermion is obtained if we introduce a background gauge field to cancel the mass of fermions 
which arise from the spin connection for the positive curvature of $S^2$.
The Lagrangian of our model is then constructed by specifying the gauge symmetry, the field contents and 
the boundary conditions for each fields.
The gauge symmetry is chosen as the SM gauge symmetry with the extra U(1)$_X$ symmetry, which is 
SU(3)$\times$SU(2)$\times$U(1)$_Y \times$U(1)$_X$, where 
the extra U(1)$_X$ symmetry is introduced so that all the fermions in our model have the massless modes corresponding to 
the SM fermions.
We then introduced the field contents where the zero modes of the fields correspond to the SM field contents under 
their boundary conditions.
Thus the combinations of chirality and boundary conditions for each fermions are determined
 to give the zero mode
which correspond to the SM fermions as summarized in Table.~\ref{conditions}.

We then analyzed the KK mode expansion for fermions, gauge fields,
 and Higgs field.
The fermions are expanded in terms of the linear combinations of the eigenfunctions of the extra-space Dirac operator 
which contains background gauge field.
Those linear combinations are defined
 to satisfy the boundary conditions of the fermions.
After the mode expansion and integrating $S^2$ coordinates, we obtained the kinetic term and the KK mass term of fermions 
in four-dimensional spacetime,
and
 confirmed that
 each fermions have the chiral massless mode.
The mass spectrum of the fermion KK mode is then obtained as in Eqs.~(\ref{massf}),(\ref{degeneracy1}) and (\ref{degeneracy2}).
The gauge fields are expanded in terms of the linear combinations of the spherical harmonics which satisfy
the boundary condition.
There we used substitution Eq.~(\ref{okikae1}) and (\ref{okikae2}) for the extra components of gauge field 
to diagonalize KK mass term of them. 
We then specified the NG boson component originated from extra components of gauge fields
 which will be eaten by KK modes of four dimensional components of gauge field giving their 
longitudinal component.
We then obtained the quadratic terms of the gauge fields in four-dimensional spacetime as in Eq.~(\ref{4Dqdrtc}), 
after gauge fixing and integration of the $S^2$ coordinates.
We then analyzed the mass spectrum of the gauge fields and summarized the feature of the mass spectrum in Eq.~(\ref{massg}) and in
the sentences below Eq.~(\ref{massg}).
There we noted that the U(1)$_X$ symmetry is anomalous and is broken at the quantum level, so that 
its gauge boson should be heavy. 
We thus expect the U(1)$_X$ gauge boson and its KK modes are decoupled from the low energy sector of our model. 
The Higgs field is also expanded in terms of the linear combinations of the spherical harmonics which satisfy
the boundary condition.
The mass spectrum of the Higgs KK modes is specified in Eq.~(\ref{massh}) and sentences below Eq.~(\ref{massh}).
These mass spectrum are summarized in Table
~\ref{mass-spectrum}

We also investigated the KK-parity in our model and found that the KK-parity is defined as $(-1)^m$.
This KK-parity is conserved as a 
result of $Z_2'$ symmetry of the Lagrangian and 
when the stability of the
lightest KK particle with $m$=odd is confirmed.
We, therefore, found that the lightest KK photon $\gamma_{\mu}^{11}$, which is the promising candidate of the lightest KK particle, is 
stable and can be a good candidate of the dark matter.
We must take into account the quantum correction to the masses of the KK particles in order to clarify the lightest KK particle and 
the dark matter candidate.
Furthermore we need to derive all interaction terms.
However this is beyond the scope of this paper and we leave this for
 future work.

It would be very interesting to study experimental signatures of our model
and compare them with other extra dimensional model 
predictions.

\subsection*{Acknowledgments}
The work of N.~M. was supported in part by the Grant-in-Aid for Scientific Research 
of the Ministry of Education, Science and Culture, No.18204024. %
The work of T.~N. was supported in part by the Grant-in-Aid for the Ministry
of Education, Culture, Sports, Science, and Technology, Government of
Japan (No. 19010485). 
The work of J.~S. was supported in part by the Grant-in-Aid for the Ministry
of Education, Culture, Sports, Science, and Technology, Government of
Japan (No. 20025001, 20039001, and 20540251).
The work of M.~Y. was supported in part by the Grant-in-Aid for the Ministry
of Education, Culture, Sports, Science, and Technology, Government of
Japan (No. 20007555). 

\appendix

\renewcommand{\theequation}{\Alph{section}.\arabic{equation}}

\section{Geometrical quantity on $S^2$}
We summarize the geometrical quantity on $S^2$ such as vielveins $e^a_{\alpha}$, killing vectors $\xi^{\alpha}_a$
and spin connection $R_{\alpha}^{ab}$.
The vielveins are expressed as 
\begin{align}
e^1_{\theta} &= R, \nonumber \\
e^2_{\phi} &= R \sin \theta, \nonumber \\
e^1_{\phi} &= e^2_{\theta} = 0. 
\end{align}
The non-zero components of spin connection are 
\begin{equation}
R^{12}_{\phi} = - R^{21}_{\phi} = -\cos \theta.
\end{equation}



\section{Summary of the Jacobi polynomial}
We summarize the feature of the Jacobi polynomial $P_n^{\alpha,\beta}(z)$ ($\alpha,\beta>-1$) \cite{A.A.Abrikosov}.
The Jacobi polynomial $P_n^{\alpha,\beta}(z)$ obey the differential equation of the form 

\begin{equation}
\label{JacobiP1}
\sigma(y)P''+\tau(y)P'+\lambda_n P 
= \frac{1}{\rho^{(\alpha,\beta)}(y)} \frac{d}{dy}[\sigma(y) \rho^{(\alpha,\beta)}(y) P'] + \lambda_n P
=0,
\end{equation}
where $\rho^{\alpha,\beta}$, $\sigma(z)$, $\tau(z)$ and $\lambda_n$ are given by
\begin{align}
\label{JacobiP2}
\rho^{(\alpha,\beta)}(z) &= (1-z)^{\alpha}(1+z)^{\beta}, \\
\sigma(y) &= 1-z^2, \\
\tau(z) &= \beta-\alpha-(\alpha+\beta+2)z, \\
\lambda_n &= n(n+\alpha+\beta+1),
\end{align}
where
 $n$ is a non-negative integer.

The explicit form of the Jacobi polynomials are given by both the differential and integral of 
Rodrigues' formulas

\begin{align}
P_n^{(\alpha, \beta)}(z) &= \frac{(-1)^n}{2^n n!} \frac{1}{\rho^{(\alpha,\beta)}(z)} 
                           \frac{d^n}{dz^n}[\sigma(z)^n \rho^{(\alpha, \beta)}(z)], \\
P_n^{(\alpha, \beta)}(z) &= \frac{(-1)^n}{2^n n!} \frac{1}{\rho^{(\alpha,\beta)}(z)} 
                           \frac{n!}{2 \pi i} \oint \frac{\sigma^n(z) \rho^{(\alpha,\beta)}(w)}{(w-z)^{n+1}} dw,
\end{align}
where the contour of complex integration in the second equation must encircle the point $z$.
The orthogonal relation of the Jacobi polynomials is given as
\begin{equation}
\int^1_{-1} P_m^{(\alpha,\beta)}(z) P_n^{(\alpha,\beta)}(z) \rho^{(\alpha,\beta)}(z) dz 
= \delta_{mn} \frac{2^{\alpha+\beta+1} \Gamma(n+\alpha+1) \Gamma(n+\beta+1)}{n!(2n+\alpha+\beta+1)\Gamma(n+\alpha+\beta+1)}.  
\end{equation}

The recurrence equations for the Jacobi polynomials are summarized such that
\begin{align}
\label{zenka1}
\frac{d}{dy} P_n^{(\alpha,\beta)}(y) &= \frac{1}{2}(n+\alpha+\beta+1) P_{n-1}^{(\alpha+1,\beta+1)}(y)  \qquad (n>0), \\
\label{zenka2}
(\frac{\alpha}{1-y}-\frac{d}{dy}) P_n^{(\alpha,\beta)}(y) &= \frac{n+\alpha}{1-y} P_n^{(\alpha-1,\beta+1)}(y), \\
\label{zenka3}
(\frac{\beta}{1+y}+\frac{d}{dy}) P_n^{(\alpha,\beta)}(y) &= \frac{n+\beta}{1+y} P_n^{(\alpha+1,\beta-1)}(y).
\end{align}


\end{document}